\documentclass[11pt,manuscript]{aastex}
\usepackage{stmaryrd}
\usepackage{amsmath}
\usepackage{amssymb}
\usepackage{array, color}
\usepackage{tabularx}
\usepackage{threeparttable}
\usepackage{titletoc}
\usepackage{booktabs}
\usepackage[colorlinks,linkcolor=blue,anchorcolor=blue,citecolor=blue,urlcolor=blue,dvipdfm]{hyperref}

\begin{document}

\title{Quantum Langevin molecular dynamics
determination of the solar-interior equation of state}

\author{Jiayu Dai, Yong Hou, and Jianmin Yuan\altaffilmark{*}}

\affil{Department of Physics, College of Science, National
University of Defense Technology, Changsha 410073, P. R. China}

\altaffiltext{*}{Corresponding author, E-mail:jmyuan@nudt.edu.cn}

\begin{abstract}
The equation of state (EOS) of the solar interior is accurately and
smoothly determined from \textit{ab initio} simulations named
quantum Langevin molecular dynamics (QLMD) in the pressure range of
$58 \leq P \leq 4.6\times10^5$ Mbar at the temperature range of $1
\leq T \leq 1500$ eV. The central pressure is calculated, and
compared with other models. The effect of heavy elements such as
carbon and oxygen on the EOS is also discussed.
\end{abstract}


\section{Introduction}

For solar and stellar models, a high-quality equation of state (EOS)
is very crucial \citep{heli}. It is well-known that some
requirements should be satisfied for solar and stellar modeling:
thermodynamic quantities would be smooth, consistent, valid over a
large range of temperature and density, and incorporate the most
important astrophysically relevant chemical elements \citep{rev1} as
well. There are two major efforts for a high-precision and
high-accuracy EOS made before, which have been included in the
recent opacity recalculations. They are the international Opacity
Project (OP) \citep{op1,op2} and Opacity Project at Livermore
(OPAL). In OP, the model named Mihalas-Hummer-D\"{a}ppen (MHD)
\citep{mhd1,mhd2,mhd3,mhd4,mhd5} for EOS is developed, dealing with
\textit{heuristic} concepts about the modification of atoms and ions
in a plasma. In the other one, OPAL, the EOS relies on
\textit{physical picture}, which is built on the modeling of
electrons and nuclei. This model is called ACTEX (activity
expansion) EOS \citep{act1,act2,act3,act4}. Both MHD and OPAL are
dependent on the potentials between particles (electron-electron,
ion-electron, ion-ion). In addition, there are other models such as
\cite{eff} (EFF), which is thermodynamically consistent and
qualitatively correct. EFF does not consider excited states, H
molecule formation, or Coulomb corrections, and treats full
ionization for heavy elements at high density. It is interesting to
note that the effect of partially degenerate electrons can be
included partly according to the Fermi-Dirac statistics. In most
cases, the EOS from OP, OPAL and even EFF should be good for the
input parameters of stellar models. However, there is always some
physics such as the coupling of ions which is missed in these
models. With the increasing requirements of the high-precision in
helioseismic study, a parameter-free model beyond Debye-H\"{u}ckel
approximation for a more accurate EOS than that of OPAL and MHD is
still necessary \citep{rev1,dybe}.

The conditions in the solar interior are very complicated, where the
densities are from 10 g/cm$^3$ up to 160 g/cm$^3$, and temperatures
from 50 eV to 1400 eV \citep{standard}. At the same time, Hydrogen
and Helium are the main elements, taking about 98\%, with small
abundances of other heavy elements such as carbon, oxygen and iron.
In order to obtain an accurate and smooth EOS for the whole sun, it
is necessary to develop a model which can cover all conditions in
the sun. It is very clear that matter in the sun is not always ideal
ionized gas plasma, but moderately coupled, partly degenerate, and
partly ionized (especially for heavy elements) in some area
according to the definition of coupling parameter $\Gamma$ and
degenerate parameter $\theta$ \citep{str}, where
$\Gamma=Z^{*2}/(k_BTa)$, with T the system temperature, $k_B$ the
Boltzmann constant, $a$ the mean ionic sphere radius defined as
$a=(3/(4{\pi}n_i))^{1/3}$, $Z^*$ the average ionization degree,
$n_i$ the ionic number density, and $\theta=T/T_F$, with the Fermi
temperature $T_F=(3\pi^2n_e)^{2/3}/2$ ($n_e$ is the number density
of electrons). Generally speaking, when the degenerate parameter
$\theta \sim 1$ and coupled parameter $\Gamma \sim 1$, the matter is
considered to be partially degenerate and moderately coupled.
Otherwise, if $\theta \ll 1$ ($\theta \gg 1$) and $\Gamma \gg 1$
($\Gamma \ll 1$), matter is strongly (weakly) degenerate and
strongly (weakly) coupled. For most of regimes in solar interior,
matter is weakly coupled and weakly degenerate. However, the
extremely high density can not promise the negligibility of the
coupling of ions, and the theory of ideal ionized gas plasma is not
always appropriate. For weakly coupled or weakly degenerate matter,
single atomic models such as average atoms (AA) models
\citep{AA,aamix}, and detailed level accounting (DLA) \citep{dla}
are assumed valid. Generally, AA model is built for the dense
plasma, and DLA model is available for the relatively low density.
Very recently, AA model was applied for the astrophysical plasmas
successfully \citep{aanew}. However, there is no direct physical
evidence supporting these approaches, and we don't even know whether
they are correct in such dense matter, and how they behave for the
extreme dense matter under conditions such as the solar center.
Therefore, it is extremely important to develop a more reliable
model for all these conditions.

For the strongly (moderately) coupled matter with partly degenerate
electrons, quantum molecular dynamics (QMD), without requiring any
assumptions about the potential between atoms, supplies a powerful
and accurate tool, and has been successfully applied in warm dense
matter (WDM) \citep{qmd,qmd1,qmd2,qmd3}. In astrophysics, QMD has
been used to study the properties of giant planets and given rise to
amazingly satisfying results
\citep{astro1,astro2,astro3,astro4,astro5,astro6}. Very recently,
QMD was extended to the field of high energy density physics (HEDP)
by considering the electron-ion collision induced friction (EI-CIF)
in the procedure of molecular dynamics, and the corresponding model
called quantum Langevin molecular dynamics (QLMD) was constructed
\citep{lg,qlmd}. Thanks to this model, one can go into the regime of
HEDP from first principles, and the thermodynamic properties of the
sun and sun-like stars can be therefore researched based on
\textit{ab initio} method. It also gives us a good tool to
investigate the system in which the coupling of ions is important.

In this work, we firstly study the EOS of conditions under the solar
interior using QLMD. Three densities with different temperatures are
chosen, and their EOS are compared with the EOS of the AA model with
energy-level broadening (AAB) \citep{AA}. QLMD can ensure much more
accurate results in all conditions, and agrees with AAB at high
temperatures. The central pressure of the sun is also calculated
using different models, comparing with the data of other models and
standard solar models \citep{standard}. The important effect of
heavy elements on EOS of very dense matter is also investigated at
the end.

\section{Theoretical method}

\subsection{Quantum Langevin molecular dynamics}

We firstly recall the QLMD method here, where the electronic
structure is studied from density functional theory (DFT), and the
ionic trajectory is performed using Langevin Equation \citep{qlmd}
\begin{equation}
\label{e1}
  M_I\mathbf{\ddot{R}_I}=\mathbf{F}-{\gamma}M_I\mathbf{\dot{R}_I}+\mathbf{N_I},
\end{equation}
Where $M_I$ is the ionic mass, $\mathbf{F}$ is the force calculated
in DFT, $\gamma$ is a Langevin friction coefficient, $\mathbf{R_I}$
is the position of ions, and $\mathbf{N_I}$ is a Gaussian random
noise corresponding to $\gamma$. Considering the dynamical
collisions between electrons and ions at high temperature, the
friction coefficient $\gamma$ can be estimated by
\begin{equation}
\label{e2}
\gamma=2{\pi}\frac{m_e}{M_I}Z^*(\frac{4{\pi}n_i}{3})^{1/3}\sqrt{\frac{k_BT}{m_e}}
\end{equation}
Where $m_e$ is the electronic mass, $Z^*$, $n_i$ and T are the same
as the above definition. Therefore, the dynamical electron-ion
collisions at high temperature can be described as a friction or
noise effect. By introducing this dynamical EI-CIF, the first
principles simulations can be applied in the HEDP regime, overcoming
the difficulty of numerical calculations, and therefore give more
reliable results.

To guarantee an accurate sampling of the Maxwell-Boltzmann
distribution, the noise has to obey the fluctuation-dissipation
theorem:
\begin{equation}
\label{e3}
  <\mathbf{N_I}(0)\mathbf{N_I}(t)>=6\gamma M_Ik_BTdt
\end{equation}
Where $dt$ is the time step of molecular dynamics. At the same time,
the random forces are taken from a Gaussian distribution of mean
zero and variance of $<\mathbf{N_I}^2>=6\gamma M_Ik_BT/dt$.

In order to integrate Eq.~\ref{e1}, the formalism in a Verlet-like
form \citep{Pastor} integration is performed as follows
\begin{equation}
\label{e4}
\begin{split}
  \mathbf{R_I}(t+dt)&=\mathbf{R_I}(t)+(\mathbf{R_I}(t)-\mathbf{R_I}(t-dt))\frac{1-\frac{1}{2}\gamma_Tdt}{1+\frac{1}{2}\gamma_Tdt}\\
                           &+(dt^2/M_I)(\mathbf{F_{BO}}(t)+\mathbf{N_I}(t)(1+\frac{1}{2}\gamma_Tdt)^{-1},
\end{split}
\end{equation}

The velocities of ions $\mathbf{v_I}(t+dt)$ can also be calculated
by using the Verlet formula
\begin{equation}
\label{e5}
 \mathbf{v_I}(t+dt)=\mathbf{\dot{R}_I}=\frac{\mathbf{R_I}(t+dt)-\mathbf{R_I}(t-dt)}{2dt}.
\end{equation}

This \textit{ab initio} molecular dynamics model based on LE is
named QLMD, which extends the applications of \textit{ab initio}
method into the field of HEDP. Based on QLMD, the effects of coupled
ions and degenerate electrons can be studied, and give much more
accurate results at relatively low temperature and high density. It
is worth pointing out that the computational cost of the QLMD is
very expensive under conditions of very weakly coupled and
non-degenerate matter. In this case, the existing models such as
MHD, ACTEX, and AA model can give consistent and accurate data.

\subsection{Average atoms model with electronic energy-level broadening}

The average atom model is one of the statistical approximations
applied to study the electronic structure of atoms and ions in hot
dense plasmas, which is easily to be applied in conjunction with a
variety of treatments of electron orbitals in atoms. In a full
relativistic self-consistent field-based AA model, the influence of
the environment on the atom is assumed to be spherically symmetric
on average. The movement of an electron under the interactions of
the nucleus and other electrons is approximated by a central field,
which is determined by the standard self-consistent calculation. In
the central field, the radial part of the Dirac equation has the
form:
\begin{equation}
\label{e6}
\left\{\begin{array}{l} \frac {dP_{n\kappa}(r)}{dr}+\frac
{\kappa}{r}P_{n\kappa}(r)=
\frac {1}{c}[\epsilon+c^2-V(r)]Q_{n\kappa}(r)\\
\frac {dQ_{n\kappa}(r)}{dr}-\frac {\kappa}{r}Q_{n\kappa}(r)= -\frac
{1}{c}[\epsilon-c^2-V(r)]P_{n\kappa}(r)
\end{array}
\right.
\end{equation}
where $P(r)$ and $Q(r)$ are respectively the large and small
components of the wave function, $c$ is the light speed, and $V(r)$
is the self-consistent potential, consisting of three parts, which
are respectively the static, exchange and correlation potentials.
The static part is calculated from the charge distributions in the
atom, while the exchange and correlation parts take the approximate
forms of Dharma-Wardana and Taylor \citep{exchange}. For bound
states, we have the boundary conditions satisfied by the radial wave
functions
\begin{equation}
\label{e7}
\left\{\begin{array}{l}
P_{n\kappa}(r)\stackrel{r\rightarrow 0}{\longrightarrow} ar^{l+1}\\
P_{n\kappa}(R_b)=0
\end{array}
\right.
or
\left\{\begin{array}{l}
P_{n\kappa}(r)\stackrel{r\rightarrow 0}{\longrightarrow} ar^{l+1}\\
\frac{d}{dr}[\frac{P_{n\kappa}(r)}{r}]_{R_b}=0
\end{array}
\right.
\end{equation}
where $R_b$ is the radius of the atom. The electron distribution is
calculated separately for the bound and free electron parts. The
bound electron density is obtained according to
\begin{equation}
\label{e8}
D_b(r)=\frac {1}{4\pi r^2}\sum\limits_{j}b_{j}
(P_{j}^{2}(r)+Q_{j}^{2}(r))
\end{equation}
where $b_j$ is the occupation number of the state $j$. In AA model
without energy-level broadening, the occupation number $b_j$ is
determined by the Fermi-Dirac distribution
\begin{equation}
\label{e9}
b_{j}=\frac {2|\kappa _j|}{exp((\epsilon _j-\mu)/T)+1}.
\end{equation}
In order to consider the effect of energy-level broadening, Gaussian
functions $\rho(\epsilon)$ centered at the corresponding electron
orbital energies of Eq.~\ref{e7} are introduced into the Fermi-Dirac
distribution of electrons, i.e.,
\begin{equation}
\label{e10}
b_{j}(\epsilon)=\frac {2|\kappa
_j|\rho(\epsilon)}{exp((\epsilon _j-\mu)/T)+1}
\end{equation}
and
\begin{equation}
\label{e10}
D_b(r)=\frac {1}{4\pi
r^2}\sum\limits_{j}\int_{a}^{b}b_{j}(\epsilon)
(P_{j}^{2}(r)+Q_{j}^{2}(r))d\epsilon,
\end{equation}
where $1=\int_{a}^{b}\rho(\epsilon)d\epsilon$ and the two orbital
energies obtained from the boundary conditions of Eq.~\ref{e7} are
taken as the upper and lower half maximum positions of the Gaussian
form energy-level broadening.

Based on this approach, the splitting of the real energy levels
approximated by the energy-level broadening can be considered, which
makes the irregularities caused by the pressure-induced electron
ionization without energy-level broadening disappear naturally
\citep{AA,jcpAA}.

\section{Results and discussions}

\subsection{Computational details}

\begin{figure}[!tb]
\centering
\includegraphics*[width=5.0in]{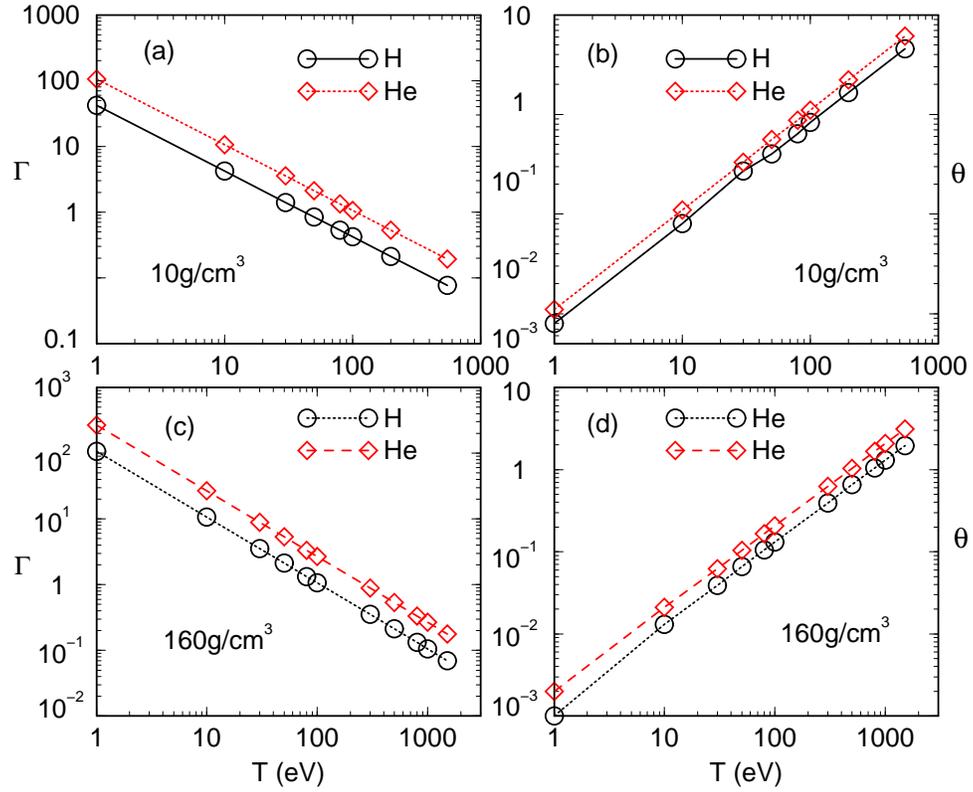}
\caption{(Color online) The coupling parameters ($\Gamma$) and
degenerate parameters ($\theta$) of H and He from 1 eV to 550eV at
10 g/cm$^3$ (a)-(b), and from 10 eV to 1500 eV at the density of 160
g/cm$^3$ (c)-(d).} \label{para}
\end{figure}

For the study of solar interior EOS, we choose three typical
densities in the solar interior: 10 g/cm$^3$ (radiative zone), 100
g/cm$^3$ and 160 g/cm$^3$ (core). Temperatures from 1 eV to 550 eV
for 10 g/cm$^3$, and 10 eV to 1500 eV for 100 and 160 g/cm$^3$ are
calculated, covering the conditions from the radiative zone into the
core. Since the H and He elements take up about 98\% of the
composition of the sun \citep{chem}, we calculated four structures
with different compositions (using X to represent the mass abundance
of H, and Y for abundance of He, Z for others), which are X=1, Y=0,
Z=0; X=0, Y=1, Z=0; X=0.7, Y=0.3, Z=0; X=0.40, Y=0.60, Z=0. The
supercells containing T=125 particles for 10 g/cm$^3$ and T=256
particles for 160 g/cm$^3$ are constructed. When Z=0, the number of
H atoms ($N_H$) and He ($N_{He}$) can be calculated by the formulas:
\begin{equation}
\label{e12} \frac{4N_{He}}{4N_{He}+N_H}=Y, N_{He}+N_H=T
\end{equation}
In these conditions, the matter goes from strongly coupled to
relatively weakly coupled, and from partially degenerate to weakly
degenerate, as shown in Fig.~\ref{para}, where QLMD has been proved
successfully.

For the QLMD simulations, we have used the Quantum Espresso package
\citep{pwscf} based on the finite temperature DFT. The Perdew-Zunger
parametrization of local density approximation (LDA) \citep{pz} is
used for the exchange-correlation potential. Similar to the
calculation of Hydrogen at the density of 80 g/cm$^3$
\citep{qlmd,hydro}, the Coulombic pseudopotentials with a cutoff
radius of 0.005 a.u. for H and He are adopted. The plane wave cutoff
energy is set to be from 200 Ry to 400 Ry with the increase of
temperature. As discussed in Ref.~\citep{hydro}, pressure
delocalization promises that the upper band electronic eigenstates
are nearly plane waves, and thus the basis set is greatly reduced.
Therefore, even though we choose a large number of bands in order to
ensure the accuracy of the calculation, the computational cost is
not very expensive. In our cases, we used enough bands in order to
make the corresponding band energies higher than 8$k_B$T (especially
at high temperature). Gamma point only is used for the
representation of the Brillouin zone. We tested all these parameters
carefully, and found that more k-points, more bands, larger energy
cutoff, and more particles do not give any significant difference.
The time step of QLMD is $a_I/(20\sqrt{k_BTM_I})$ \citep{hydro},
where $a_I$ and $M_I$ are the average ionic radius and the ionic
mass of $I_{th}$ ion, respectively. After thermalization, each
structure is simulated for 5000 to 10000 time steps to pick up the
useful information.

\subsection{Equation of state}

\begin{figure}[!tb]
\centering
\includegraphics*[width=5.0in]{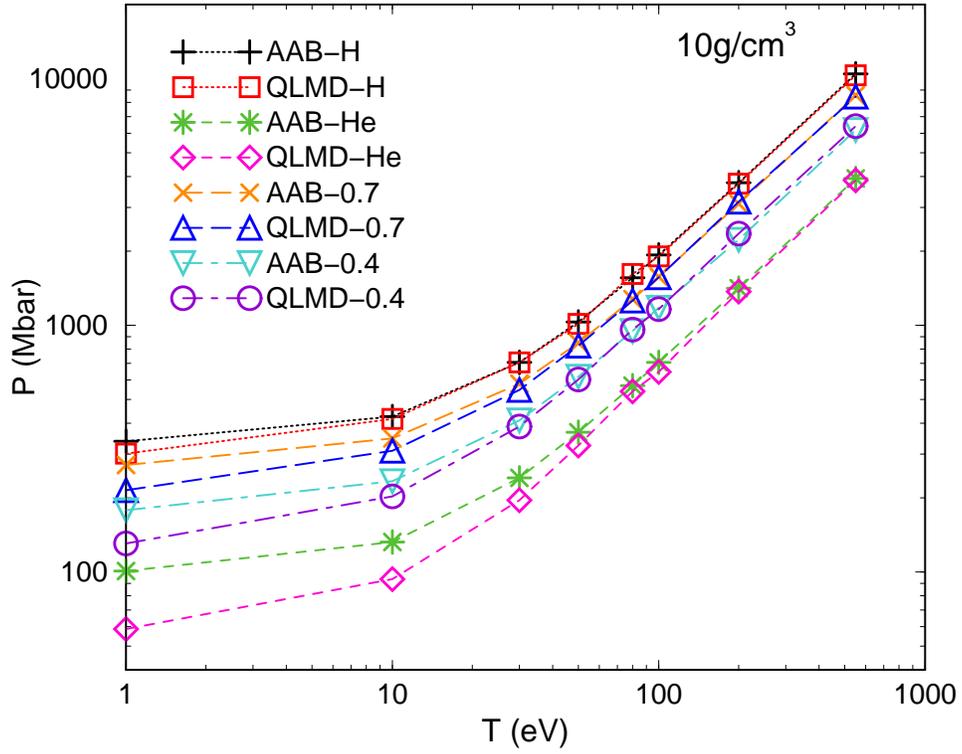}
\caption{(Color online) Pressure vs temperature for the density of
10 g/cm$^3$ with different chemical compositions compared with
results of AAB. In the figure, 0.7 represents X=0.7, 0.4 represents
X=0.4, respectively.} \label{10g}
\end{figure}

Firstly, we study the isochoric heating curve along the density of
10 g/cm$^3$ from 1 eV to 550 eV under the conditions of the solar
radiative zone. Comparisons between the pressure calculated using
QLMD and AAB models for different compositions are given in
Fig.~\ref{10g}. It is shown that for the system of He with two
electrons at lower temperatures, the pressure of AAB model is much
larger than that of QLMD. However, the relative difference for H is
not as big as for He. Therefore, with the increase of temperature
and abundance of Hydrogen, the gap between QLMD and AAB models
becomes less and less. It is very clear that more electrons and low
temperature result in difficulties of describing the ionization
balance for AAB model, where the composition of the system is
complicated. The strong coupling among the ions, which is included
in QLMD model but not in AAB model, plays an important role too in
the calculations. Here, we recall that the simple way to calculate
the EOS containing only one element and averaging the pressure
according to the mass ratio is not accurate enough, as discussed in
Ref.~\citep{aamix}. For example, for the density-temperature point
(10 g/cm$^3$, 10 eV) and composition (X=0.7, Y=0.3, Z=0), the
average pressure is $P_a=(0.7\times301.3+0.3\times58.6)=228.49$
Mbar. However, the pressure of this mixture obtained using QLMD is
214.77 Mbar, and the error is about 6\%, which is significant for
astrophysical applications.

\begin{figure}[!tb]
\centering
\includegraphics*[width=5.0in]{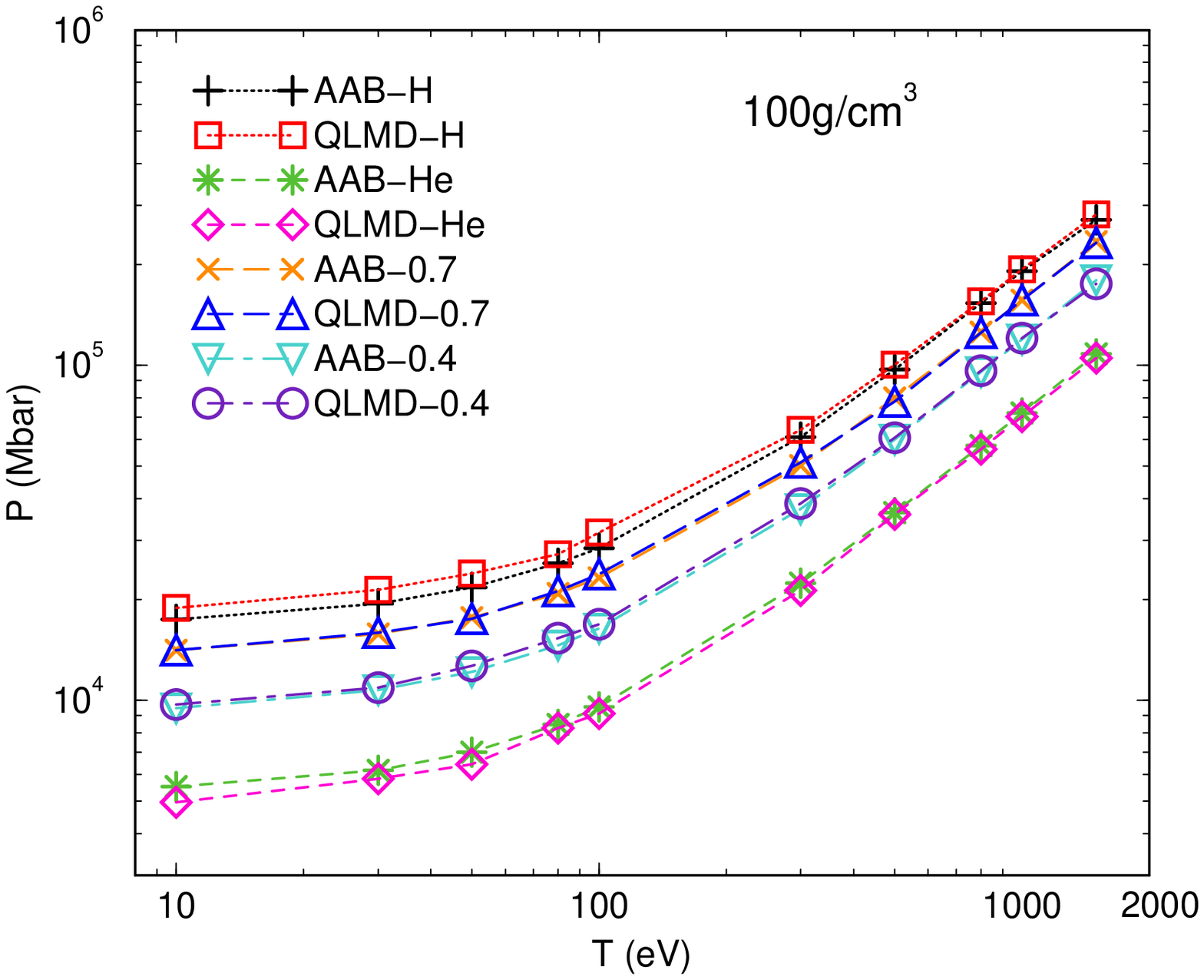}
\caption{(Color online) Pressure vs temperature for the density of
100 g/cm$^3$ with different chemical compositions compared with
results of AAB.} \label{100g}
\end{figure}

\begin{figure}[!tb]
\centering
\includegraphics*[width=5.0in]{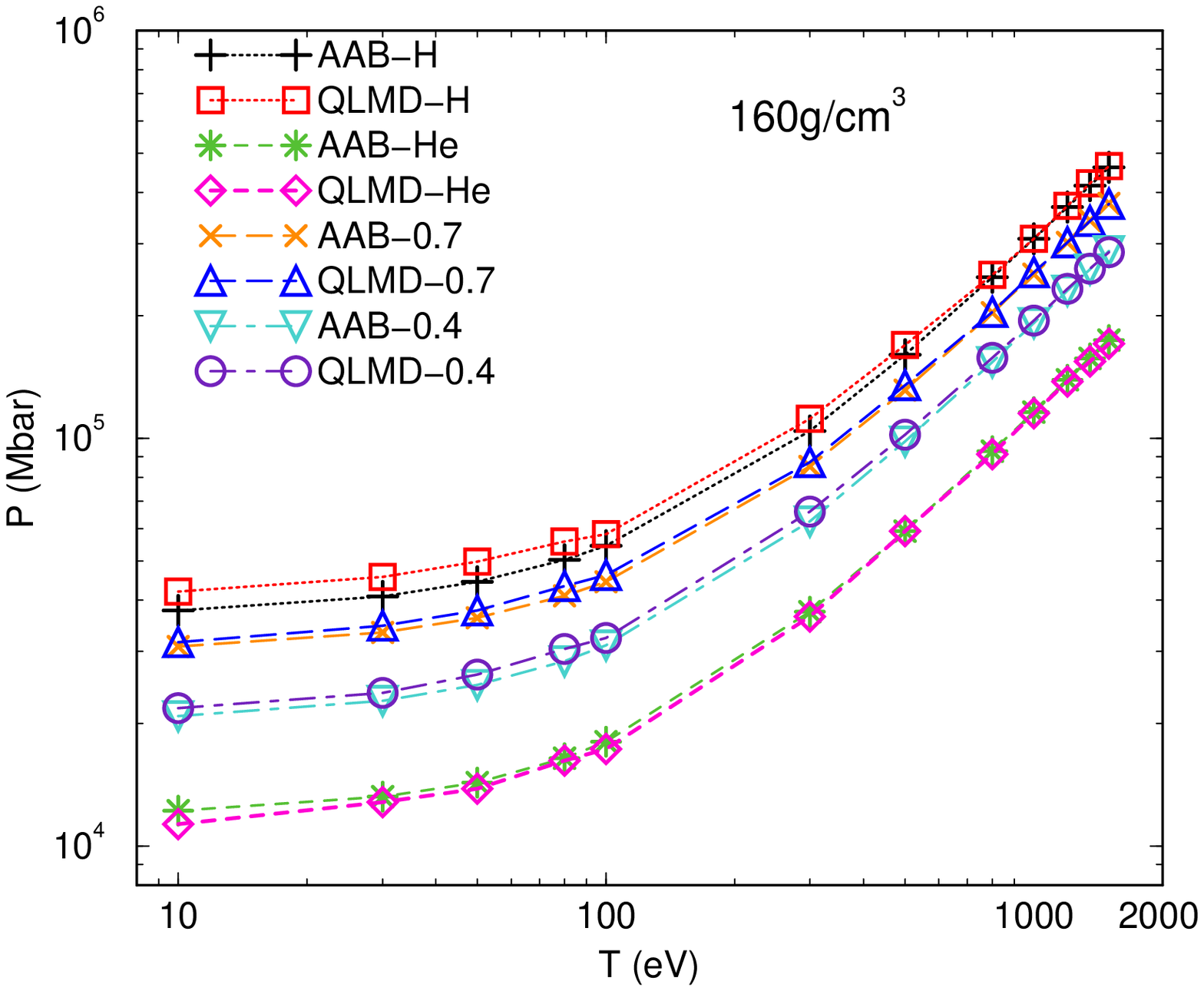}
\caption{(Color online) Pressure vs temperature for the density of
160 g/cm$^3$ with different chemical compositions compared with
results of AAB.} \label{160g}
\end{figure}

When the conditions go further into the solar interior, some
interesting characteristics appear. For the density of 100 g/cm$^3$,
pressure-temperature relation is shown in Fig.~\ref{100g}. Moreover,
the maximum density of the sun is about 160 g/cm$^3$, whose
isochoric heating curve from 10 eV to 1500 eV is shown in
Fig.~\ref{160g}. Contrary to the relative difference between QLMD
and AAB models in Fig.~\ref{10g}, the pressures of AAB model for 100
and 160 g/cm$^3$ are smaller than those of QLMD model, especially
for one component H. In fact, the spherical assumption \citep{aamix}
for ions in AAB model makes the effective volume of the system
larger than the real one, giving rise to smaller pressure. This
phenomenon is obvious when the density is high enough, such as 160
g/cm$^3$ for H here. With the increase of temperature, the pressure
of AAB and QLMD becomes consistent. At high temperatures, the ions
are weakly coupled and electrons are almost free. Therefore, the
semiclassical methods such as Thomas-Fermi and AAB can work well.
Furthermore, it can be deduced that the EOS of the whole area of the
sun can be obtained accurately and smoothly, which is beyond the OP
and OPAL models, and completely parameter-free.

For the central conditions in the sun, the density reaches up to
152.7 g/cm$^3$ and the temperature up to 15696000 K according to the
standard solar model \citep{standard}. The central pressure is also
about 2.342$\times$10$^5$ Mbar, which is never reached from first
principles simulation before. Based on QLMD, we solved this problem
from an \textit{ab initio} approach and obtained the EOS under this
condition. We firstly compared the pressure close to the solar
center with other models, as shown in Table.~\ref{cent}, and they
agree reasonably well with each other. It is interesting to find
that the pressure of AA model without energy-level broadening (AANB)
is much larger than those of QLMD and AAB, and consistent with other
models of EFF, Livermore (LIV) and MHD \citep{eos}. Since QLMD can
be realized much more reliably, we can think AAB more reasonable,
and both QLMD and AAB can improve the accuracy of EOS much. In fact,
for the very dense medium, the electronic structures of elements are
not only the energy levels, but also energy bands. It is reasonable
to consider the effect of energy-level broadening, which is
naturally included in QLMD and thus gives rise to appropriate
results. For the solar cental regime, we calculate the pressure with
different compositions. First of all, we consider the chemical
compositions of H and He, with abundances X=0.34828, Y=0.65172,
Z=0.0. It is found that the pressures of QLMD and AAB are almost
equal within 0.5\% error and are very similar to that of Standard.
The small difference is caused by the lack of the heavier elements.
In order to understand the effect of heavier elements on the EOS, we
study the pressure of chemical compositions of H, He, C and H, He,
O, respectively. As shown in Table.~\ref{cent}, the existence of
heavier elements decreases the pressure a little, since the
abundance is very small. In principle, we can calculate the
conditions of solar interior with any composition and any abundance
using QLMD resulting in EOS accurately and smoothly, particularly at
relatively low temperatures and high densities. QLMD can also be
very complementary to the AA models considering the computational
cost, and improve significantly the accuracy of EOS in the solar
interior.

In conclusion, the EOS of solar interior is calculated based on
first principles method named QLMD, which is beyond the MHD and OPAL
models. Furthermore, the model of QLMD can promise the accuracy for
the EOS and be complementary to AA-like models. The accuracy can be
improved significantly, particularly at relatively low temperatures
and high densities where the coupling of ions can not be neglected.
This gives us a useful tool to investigate the properties of the sun
and sun-like stars from \textit{ab initio} approaches, which is very
powerful in astrophysics. This work can also open a new field to
investigate the solar or stellar properties from first principles,
promoting the study in astrophysics both in theory and experiment.

\acknowledgments This work is supported by the National Natural
Science Foundation of China under Grant Nos. 10734140, 60921062 and
10676039, the National Basic Research Program of China (973 Program)
under Grant No. 2007CB815105. Calculations are carried out at the
Research Center of Supercomputing Application, NUDT. The authors
thank the comments of the anonymous reviewer.

\clearpage

\begin{table}[!tb]
\begin{center}
\caption{\label{cent} Comparison of central pressures of the sun at
typical temperature (T) and density ($\rho$). The results of EFF,
LIV (OPAL) and MHD are in Ref.~\citep{eos} (X=0.34828, Y=0.65172,
Z=0). Results of AANB, AAB, and standard solar model (Standard)
\citep{standard} are also compared. The pressures of chemical
composition of H, He and C with X=0.3387, Y=6613, Z=0 ($^a$),
X=0.3125, Y=0.6406, Z=0.0469 ($^b$), and chemical composition of H,
He and O with X=0.3077, Y=0.6308, Z=0.0615 ($^c$) are also shown.}
\begin{tabular}{lccc}
\tableline\tableline
  $\rho$ (g/cm$^3$)&       T (eV)      & EOS model   & pressure (Mbar)        \\
\hline
   141.25          &    989.45         &   EFF       & 1.6396$\times$10$^5$    \\
                   &                   &   LIV       & 1.6162$\times$10$^5$    \\
                   &                   &   MHD       & 1.6190$\times$10$^5$    \\
                   &                   &   AANB      & 1.6144$\times$10$^5$    \\
                   &                   &   AAB       & 1.5922$\times$10$^5$    \\
                   &                   &   QLMD      & 1.6090$\times$10$^5$    \\ \hline
   141.25          &    1189.58        &   EFF       & 1.9579$\times$10$^5$    \\
                   &                   &   LIV       & 1.9337$\times$10$^5$    \\
                   &                   &   MHD       & 1.9365$\times$10$^5$    \\
                   &                   &   AANB      & 1.9330$\times$10$^5$    \\
                   &                   &   AAB       & 1.9027$\times$10$^5$    \\
                   &                   &   QLMD      & 1.9122$\times$10$^5$    \\ \hline
   152.70          &    1352.64        &   Standard  & 2.3420$\times$10$^5$    \\
                   &                   &   AANB$^a$  & 2.3603$\times$10$^5$    \\
                   &                   &   AAB$^a$   & 2.3417$\times$10$^5$    \\
                   &                   &   QLMD$^a$  & 2.3525$\times$10$^5$    \\
                   &                   &   AANB$^b$  & 2.2678$\times$10$^5$    \\
                   &                   &   AAB$^b$   & 2.2534$\times$10$^5$    \\
                   &                   &   QLMD$^b$  & 2.2575$\times$10$^5$    \\
                   &                   &   AANB$^c$  & 2.2445$\times$10$^5$    \\
                   &                   &   AAB$^c$   & 2.2301$\times$10$^5$    \\
                   &                   &   QLMD$^c$  & 2.2328$\times$10$^5$    \\
\tableline
\end{tabular}
\end{center}
\end{table}

\clearpage

\clearpage


\begin{thebibliography}{}
\bibitem[Basu \& Antia(2008)]{heli} Basu, S., \& Antia, H. M. 2008, Phys. Rep. 457,
217
\bibitem[Berrington(1997)]{op2} Berrington, K. A. 1997 \textit{The Opacity Project, vol. II.}
(Bristol: Institute of Physics)
\bibitem[Collins et al.(1995)]{qmd} Collins, L. \textit{et al.}, 1995, \pre, 52, 6202
\bibitem[Dai \& Yuan(2009)]{lg} Dai, J., \& Yuan, J. 2009, Europhys. Lett. 88, 20001
\bibitem[Dai et al.(2010)]{qlmd} Dai, J., Hou, Y., \& Yuan, J. 2010, \prl, 104, 245001
\bibitem[D\"{a}ppen et al.(1988)]{mhd3} D\"{a}ppen, W., Mihalas, D., Hummer, D. G., \& Mihalas B. W. 1988, \apj, 332, 261
\bibitem[D\"{a}ppen et al.(1990)]{eos} D\"{a}ppen, W., Lebreton, Y., \& Rogers, F. 1990, Solar
Physics, 128, 35
\bibitem[D\"{a}ppen \& Nayfonov(2000)]{dybe} D\"{a}ppen, W., \& Nayfonov, A. 2000, Astrophys. J. Suppl.
Ser. 127, 287
\bibitem[D\"{a}ppen(2006)]{rev1} D\"{a}ppen, W. 2006 J. Phys. A: Math. Gen. 39, 4441
\bibitem[Desjarlais et al.(2002)]{qmd1} Desjarlais, M. P. \textit{et al.}, 2002, \pre, 66, 025401(R)
\bibitem[Dharma-Wardana \& Taylor(1981)]{exchange} Dharma-Wardana, M. W. C., \&
Taylor, R., J. Phys. C: Solid State Phys. 14, 629
\bibitem[Eggleton, Faulkner, \& Flannery(1973)]{eff} Eggleton, P., Faulkner, J., \& Flannery, B. P. 1973, Astron.
Astrophys. 23, 325
\bibitem[Faussurier et al.(2010)]{aanew} Faussurier, G., Blancard,
C., Coss\'{e}, P., \& Renaudin, P. 2010, Phys. Plasmas, 17, 052707
\bibitem[Giannozzi et al.(2009)]{pwscf} Giannozzi, P. \textit{et al.}, 2009, J. Phys.: Condens. Matter,
21, 395502; \url{www.quantum-espresso.org}
\bibitem[Gillan et al.(2006)]{astro3} Gillan, M. J., Alf\`{e}, D., Brodholt, J., Vocadlo, L., \&
Price, G. D. 2006, Rep. Prog. Phys. 69, 2365
\bibitem[Hou et al.(2006)]{AA} Hou, Y., Jin, F., \& Yuan, J. 2006, Phys. Plasmas
13, 093301
\bibitem[Hou et al.(2007)]{jcpAA} Hou, Y., Jin, F., \& Yuan, J. 2007,
J. Phys.: Condens. Matter, 19, 425204
\bibitem[Hummer \& Mihalas(1988)]{mhd1} Hummer, D. G., \& Mihalas, D. 1988, \apj, 331, 794
\bibitem[Ichimaru(1982)]{str} Ichimaru, S. 1982, Rev. Mod. Phys., 54, 1017.
\bibitem[Iglesias \& Rogers(1991)]{act4} Iglesias, C. A., \& Rogers, F. J. 1991, 371,
408
\bibitem[Iglesias \& Rogers(1993)]{act5} Iglesias, C. A., \& Rogers, F. J. 1993, \apj, 412, 752
\bibitem[Iglesias \& Rogers(1995)]{act6} Iglesias, C. A., \& Rogers, F. J. 1995, \apj, 443, 460
\bibitem[Iglesias \& Rogers(1996)]{act7} Iglesias, C. A., \& Rogers, F. J. 1996, \apj, 464, 943
\bibitem[Lodders(2003)]{chem} Lodders, K. 2003, \apj, 591, 1220
\bibitem[Lorenzen et al.(2009)]{astro1} Lorenzen, W., Holst, B. \& Redmer, R. 2009, \prl, 102, 115701
\bibitem[Mazevet et al.(2005)]{qmd2} Mazevet, S. \textit{et al.}, 2005, \pre, 71, 016409
\bibitem[Mazevet et al.(2008)]{qmd3} Mazevet, S. \textit{et al.}, 2008, \prl, 101, 155001
\bibitem[Mihalas et al.(1988)]{mhd2} Mihalas, D., D\"{a}ppen, W., \&  Hummer, D. G. 1988, \apj, 331, 815
\bibitem[Militzer \& Hubbard(2009)]{astro2} Militzer, B. \& Hubbard, W. B. 2009, Astrophys Space Sci. 322,
129
\bibitem[Bahcall et al.(2001)]{standard} Bahcall, J. N.,
Pinsonneault, M. H., \& Basu, S. 2001, \apj, 555, 990
\bibitem[Nayfonov et al.(1999)]{mhd4} Nayfonov, A., D\"{a}ppen, W., Hummer, D. G., \& Mihalas, D. M. 1999
\apj, 526, 451
\bibitem[Nettelmann et al.(2008)]{astro5} Nettelmann, N., Holst, B., Nettelmann, A., French, M., \&
Redmer, R. 2008, \apj, 683, 1217
\bibitem[Perdew \& Zunger(1981)]{pz} Perdew, J. P. \& Zunger, A. 1981, \prb, 23,
5048
\bibitem[Pastor et al.(1988)]{Pastor} Pastor, R. W., Brooks, B. R. \& Szabo,
A., 1988, Mol. Phys., 65, 1409
\bibitem[Recoules et al.(2009)]{hydro} Recoules, V., Lambert, F., Decoster, A., Canaud, B., \& Cl\'{e}rouin, J.
2009, \prl, 102, 075002
\bibitem[Rogers(1986)]{act1} Rogers, F. J. 1986, \apj, 310, 723
\bibitem[Rogers et al.(1996)]{act2} Rogers, F. J., Swenson, F. J., \& Iglesias C. A. 1996, \apj, 456, 902
\bibitem[Rogers \& Nayfonov(2002)]{act3} Rogers, F. J., \& Nayfonov, A. 2002, \apj, 576, 1064
\bibitem[Seatom(1995)]{op1} Seaton, M. J. 1995, \textit{The Opacity Project, vol. I.} (Bristol: Institute of Physics)
\bibitem[Trampedach et al.(2006)]{mhd5} Trampedach, R., D\"{a}ppen, W., \& Baturin, V. A. 2006, \apj, 646, 560
\bibitem[Vorberger et al.(2007)]{astro4} Vorberger, J., Tamblyn, I., Militzer, B., \& Bonev, S. A. 2007,
\prb, 75, 024206
\bibitem[Wilson \& Militzer(2010)]{astro6} Wilson, H. F., \& Militzer, B. 2010,
\prl 104, 121101
\bibitem[Yuan(2002)]{aamix} Yuan, J. 2002, \pre, 66, 047401
\bibitem[Zeng \& Yuan(2004)]{dla} Zeng, J., \& Yuan, J. 2004, \pre, 70,
027401
\end{thebibliography}
\end{document}